\documentclass[aps,prb,reprint]{revtex4-2}

\usepackage{graphicx}
\usepackage{amsmath}
\usepackage{comment}
\usepackage{braket}
\usepackage{xcolor}
\usepackage{hyperref}
\hypersetup{
   colorlinks=true,
   citecolor=teal,
   linkcolor=teal,
   filecolor=teal,      
   urlcolor=teal,
}
\usepackage[nameinlink,capitalise]{cleveref}

\usepackage[normalem]{ulem}

\newcommand{\be}{\begin{eqnarray}}
\newcommand{\ee}{\end{eqnarray}}

\begin{document}

%Title of paper
\title{Finite-Temperature Ferromagnetic Correlations of the Kagome Lattice Hubbard Model}% \\
%A useful experimental tool}

% List of authors 
\author{Francisco Correia}
\email[]{These authors contributed equally to this work.}
\affiliation{Department of Physics and Astronomy, San Jos\'e State University, San Jos\'e, California 95192, USA}
\author{Kyle Corbett}
\email[]{These authors contributed equally to this work.}
\affiliation{Department of Physics and Astronomy, San Jos\'e State University, San Jos\'e, California 95192, USA}
\author{Ehsan Khatami}
\email[]{ehsan.khatami@sjsu.edu}
\affiliation{Department of Physics and Astronomy, San Jos\'e State University, San Jos\'e, California 95192, USA}

\date{\today}

\begin{abstract}
The Kagome lattice Fermi-Hubbard model is one of the most physically rich, and at the same time most challenging, models to study in strongly-correlated physics. Among its special features are geometric frustration and a flat energy band that create conditions favorable to ferromagnetism near the band insulating limit. Here, we utilize two exact finite-temperature methods, the numerical linked-cluster expansion and the determinant quantum Monte Carlo, to study the extent as well as doping and interaction dependence of ferromagnetic correlations and other thermodynamic properties of the model. We find that repulsive interactions enhance ferromagnetic correlations at high electron densities and that increasing the interaction strength, extends the region with strong ferromagnetic correlations towards half filling, smoothly connecting it to Nagaoka ferromagnetism near the Mott insulating region. We also use the charge compressibility to obtain an accurate estimate for the critical interaction strength for the metal-insulator transition at half filling. These results improve our understanding of the magnetic tendencies of the model away from half filling and pave the way for further studies, including with ultracold atoms in optical lattices.
\end{abstract}

\maketitle

%%%%%%%%%%%%%%%%%%%%%%%%%%%%%%%%%%%%%%%%%%%%%%%%%%%%%%%%%%%%%%%%%%%%%%%%%%%%%%%
\section{Introduction}
%%%%%%%%%%%%%%%%%%%%%%%%%%%%%%%%%%%%%%%%%%%%%%%%%%%%%%%%%%%%%%%%%%%%%%%%%%%%%%%

Geometry plays a crucial role in the properties of strongly-correlated lattice electrons. On the one hand, the band structure dominates the physics of tight-binding models when the electronic interactions are weak. On the other hand, in the strongly-interacting region, where for example, the strength of Coulomb repulsion between electrons in the Fermi-Hubbard model ($U$) is larger than the noninteracting bandwidth, and the charge degrees of freedom are largely suppressed near commensurate filling, geometry has a substantial influence on correlations between the spin degrees of freedom. It can promote or impede antiferromagnetism (AFM), ferromagnetism (FM), or other more exotic phases of matter, such as spin liquid or chiral spin phases.

The Kagome lattice Fermi-Hubbard model (KLHM) represents a unique model in that sense, in which many geometric features as well as strong electronic correlations collide, setting the stage for extremely rich physics that in most parts of its parameter space remain unexplored. In addition to the antiferromagnetic frustration that the geometry offers due to corner-sharing triangles, the Kagome lattice has a uniquely interesting band structure. It consists of two dispersive bands, touching at a Dirac point at the corner of the Brillouin zone and having saddle points at the zone edge, resulting in van Hove singularities in the density of states at particle densities $1/3$ and $2/3$, and a third flat band (see, e.g., Fig. 1 of Ref.~\cite{yfwy-q6y9}). The latter promotes FM, as expected from the Stoner criterion. In fact, Mielke proved that the ground state of the KLHM is ferromagnetic between densities $5/3$ and $11/6$ inside the flat band for any $U>0$~\cite{Mielke_1992} (also see Ref.~\cite{PhysRevLett.69.1608}).

Numerical studies of the KLHM are relatively scarce, owing to the challenges a highly frustrated system with many competing orders poses. So far, they have often focused on commensurate fillings~\cite{PhysRevLett.100.136404,PhysRevLett.104.196401,PhysRevB.89.155141,PhysRevB.90.035118}, including half filling with an average of one electron per site, where the system undergoes a Mott metal-insulator transition in the ground state at a critical value of the interaction $U_c$~\cite{Bernhard_2007,Ohashi2006,Kuratani2007,PhysRevB.83.195127,Higa2016,Kaufmann2021,PhysRevB.104.L121118,Thereza2023}. Others have looked at the magnetic susceptibility, pairing symmetry, and spectral functions also away from half filling~\cite{PhysRevLett.95.037001,PhysRevB.105.075118,yfwy-q6y9}. The fermion `sign problem' in the determinant quantum Monte Carlo (DQMC) method~\cite{e_loh_90,p_henelius_00,v_iglovikov_15} that generally prevents one's access to low temperature phases of Hubbard models turns out to be not as severe at this density as in some other density regions~\cite{v_iglovikov_15,yfwy-q6y9}.

In the meantime, optical lattice experiments with ultracold fermionic atoms on a closely-related frustrated geometry, the triangular lattice~\cite{m_lebrat_24,Prichard2024}, have put a spotlight on {\it kinetic} magnetism in the Fermi-Hubbard model when extremely large interactions are present in the system near the Mott phase. While Nagaoka famously proved the existence of this FM for the ground state of a bipartite lattice with $U=\infty$ and in the presence of a single hole~\cite{y_nagaoka_65}, these experiments demonstrated the finite-temperature signatures and the doping dependence of the Nagaoka FM and also Haerter-Shastry AFM~\cite{Haerter05} that arise in the particle- and hole-doped sides of the Mott insulator on the triangular lattice, respectively. In those regions, the system finds it more energetically favorable to pave the way for dopants to lower the kinetic energy as superexchange becomes prohibitively expensive. 

While similar magnetic tendencies are expected for the KLHM, whose bosonic counterpart has also been realized in optical lattices~\cite{PhysRevLett.108.045305,PhysRevLett.125.133001,PhysRevA.101.011601}, the extent of the Nagaoka FM region away from half filling, and perhaps most interestingly, how it connects to Mielke's flat band FM near the band insulator~\cite{10.1143/PTP.99.489}, as well as the dependence of relevant magnetic correlations on $U$ and density at finite temperatures are largely unknown. Only recently, the uniform susceptibility of the model in the weak to intermediate coupling regions was studies at densities up to 1.5 using DQMC, showing a pronounced growth with lowering the temperature inside the flat band~\cite{yfwy-q6y9}.

Here, we utilize two state-of-the-art complementary numerical methods, the DQMC and the numerical linked-cluster expansion~\cite{M_rigol_06,b_tang_13b}, to study finite-temperature signatures of FM in the KLHM for a wide range of interaction strengths, relevant to both flat band and Nagaoka FM, and across all particle dopings. We compute the densities of state, spin correlations at different lengths, and the static magnetic and charge susceptibilities. 

We find that short-range FM correlations develop at low temperatures near the band insulating limit of the model, even for the smallest interaction strengths we have considered. They grow larger and extend to regions with smaller and smaller dopings as $U$ increases, eventually approaching half filling, consistent with the Nagaoka limit. The results, including extrapolations of the magnetic susceptibility, suggest a Stoner-like FM ground state that starts from the band insulator and spreads to other densities in the electron-doped side as its nature changes to that of a Nagaoka ferromagnet near the Mott insulator. Using results for the charge compressibility, we are also able to obtain one of the most accurate estimates for $U_c$ of the metal-insulator transition at half filling so far. Our exact results shed light on the intriguing properties of the KLHM, especially FM tendencies of the model away from half filling.

The organization of the paper is as follows. In Sec.~\ref{sec:model}, we introduce the Hamiltonian. In Sec.~\ref{sec:methods}, we briefly describe the two numerical methods used in this study. Section~\ref{sec:results} contains our numerical results and analysis of the magnetic correlations, and Sec.~\ref{sec:discuss} presents a summary and a discussion of future directions. Finally, in the Appendix, we show results for the charge susceptibility and the estimation of $U_c$.

%%%%%%%%%%%%%%%%%%%%%%%%%%%%%%%%%%%%%%%%%%%%%%%%%%%%%%%%%%%%%%%%%%%%%%%%%%%%%%%
\section{The Model}
\label{sec:model}
%%%%%%%%%%%%%%%%%%%%%%%%%%%%%%%%%%%%%%%%%%%%%%%%%%%%%%%%%%%%%%%%%%%%%%%%%%%%%%%
We work with the Fermi-Hubbard model, whose Hamiltonian is expressed as 
\begin{equation}
H=-t\sum_{\left<ij\right> \sigma}(c^{\dagger}_{i\sigma}
c^{\phantom{\dagger}}_{j\sigma} + \textrm{h.c.}) + U\sum_i n_{i\uparrow}n_{i\downarrow}
-\mu\sum_{i\sigma} n_{i\sigma},
\label{eq:H}
\end{equation}
where $c^{\phantom{\dagger}}_{i\sigma}$ ($c^{\dagger}_{i\sigma}$) annihilates (creates) a fermion with spin $\sigma$ on site $i$, $n_{i\sigma}=c^{\dagger}_{i\sigma} c^{\phantom{\dagger}}_{i\sigma}$ is the number operator, $U$ is the onsite Coulomb interaction, and $t$ is the nearest-neighbor hopping amplitude, which is set to unity to serve as the unit of energy throughout the paper. We also work in units where $k_{\rm B}=\hbar=1$.

%%%%%%%%%%%%%%%%%%%%%%%%%%%%%%%%%%%%%%%%%%%%%%%%%%%%%%%%%%%%%%%%%%%%%%%%%%%%%%%
\section{Methods}
\label{sec:methods}
%%%%%%%%%%%%%%%%%%%%%%%%%%%%%%%%%%%%%%%%%%%%%%%%%%%%%%%%%%%%%%%%%%%%%%%%%%%%%%%

\subsection{Numerical linked-cluster expansion}
Linked-cluster expansions (LCEs)~\cite{Domb,linked} are based on the idea that any extensive property of a lattice model can be determined in the thermodynamic limit by representing the property per site as a sum of contribution from all the clusters, $c$, that can be embedded in the lattice. This sum can be represented as 
\begin{equation}
    P(\mathcal{L})/N = \sum_cL(c)\times W_p(c),
    \label{eq:LCE}
\end{equation}
where $L(c)$ is the embedding factor of the cluster (the number of ways per site it can be embedded in the lattice) and $W_p(c)$ is the weight or the {\it reduced} property for cluster $c$. This weight is calculated through the inclusion-exclusion principle by computing the property for the cluster, $P(c)$, and subtracting from it the weights (for the same property) of all the smaller clusters, $s$, that can be embedded in $c$ ({\it sub-clusters} of $c$):
\begin{equation}
    W(c)=P(c)-\sum_{s\subset c}W(s).
\end{equation}
This is a recursive process which often starts with $W(c)=P(c)$ for $c$ being a single site. LCEs [Eq.~(\ref{eq:LCE})] allow for the calculation of extensive properties in the thermodynamic limit by means of calculating those same properties on finite clusters, up to a certain size, that make up the infinite lattice. In the numerical linked-cluster expansion (NLCE)~\cite{M_rigol_06,b_tang_13b}, the latter are computed using exact numerical methods, which in this study is simply the exact diagonalization.

Since the infinite series in Eq.~(\ref{eq:LCE}) has to be truncated in reality, the convergence is generally not guaranteed. In practice, for finite-temperature studies, the series converges at high temperatures (much larger than the energy scales of the system) and continues to remain convergent at low temperatures so long as the correlations in the system do not grow larger than the scale of the largest clusters considered in the series. In fact, one can show that the NLCE is equivalent to a high-temperature series expansion with the added advantage that properties of finite clusters are computed to all powers of the inverse temperature.

We utilize a newly developed package~\cite{NLCEPack} to generate all the topologically distinct clusters with up to 9 sites on the Kagome lattice using a site expansion scheme. In the latter, the order of the expansion refers to the largest size (number of sites) up to which contributions from all clusters have been included in the series. 

To extend to convergence of the series to lower temperatures than what the bare series in Eq.~(\ref{eq:LCE}) offers, we utilize numerical resummations. Specifically, we adopt the Euler resummation~\cite{Euler,M_rigol_07a} for the last five orders of the expansion as it yields the best results in terms of convergence across all densities. In our plots, unless specified otherwise, the 8th and 9th order NLCE results are shown as dashed and solid lines, respectively. Their agreement indicates convergence and the validity of the results in the thermodynamic limit. Extensive studies and comparisons of properties to exact methods, including the DQMC, on various geometries over the years have confirmed the reliability of such resummation schemes (see Refs.~\cite{b_tang_13,e_khatami_15} for some early examples).

\subsection{Determinant quantum Monte Carlo}

We use SmoQyDQMC.jl \cite{SmoQyDQMC.jl} to implement unbiased and numerically exact determinant quantum Monte Carlo simulations of the KLHM. We use a $6 \times 6$ grid of unit cells along the lattice unit vectors, for a total of 108 sites, with periodic boundaries.
We obtain results by performing 20 independent simulations with different random number seeds for each set of parameters, each with a total of 10,000 sweeps, split into 5000 warm-up and 5000 measurement sweeps. We use bin averaging for calculating the mean and standard error of the mean. We eliminate the systematic Trotterization error associated with having a finite imaginary time step by performing separate simulations with steps $\Delta\tau=0.1, 0.05, 0.02$ and then using a linear 3-point extrapolation of quantities vs $\Delta\tau^2$ to obtain an approximation for them in the limit $\Delta\tau\to 0$. We find that this is an important step in achieving an agreement, within the DQMC error bars, with the NLCE results, especially for quantities such as spin correlation, compressibility, and magnetic susceptibility.

For temperature sweeps at a fixed density, we use MuTuner.jl \cite{MuTuner.jl} which allows us to determine the optimal value of $\mu$ to reach our target density at a given temperature. For regions where NLCE is converged, we are able to use those results as a starting point in order to speed up convergence to an optimal $\mu$.

%%%%%%%%%%%%%%%%%%%%%%%%%%%%%%%%%%%%%%%%%%%%%%%%%%%%%%%%%%%%%%%%%%%%%%%%%%%%%%%
\section{Results}
\label{sec:results}
%%%%%%%%%%%%%%%%%%%%%%%%%%%%%%%%%%%%%%%%%%%%%%%%%%%%%%%%%%%%%%%%%%%%%%%%%%%%%%%

\begin{figure}[t]
    \includegraphics[width=\linewidth]{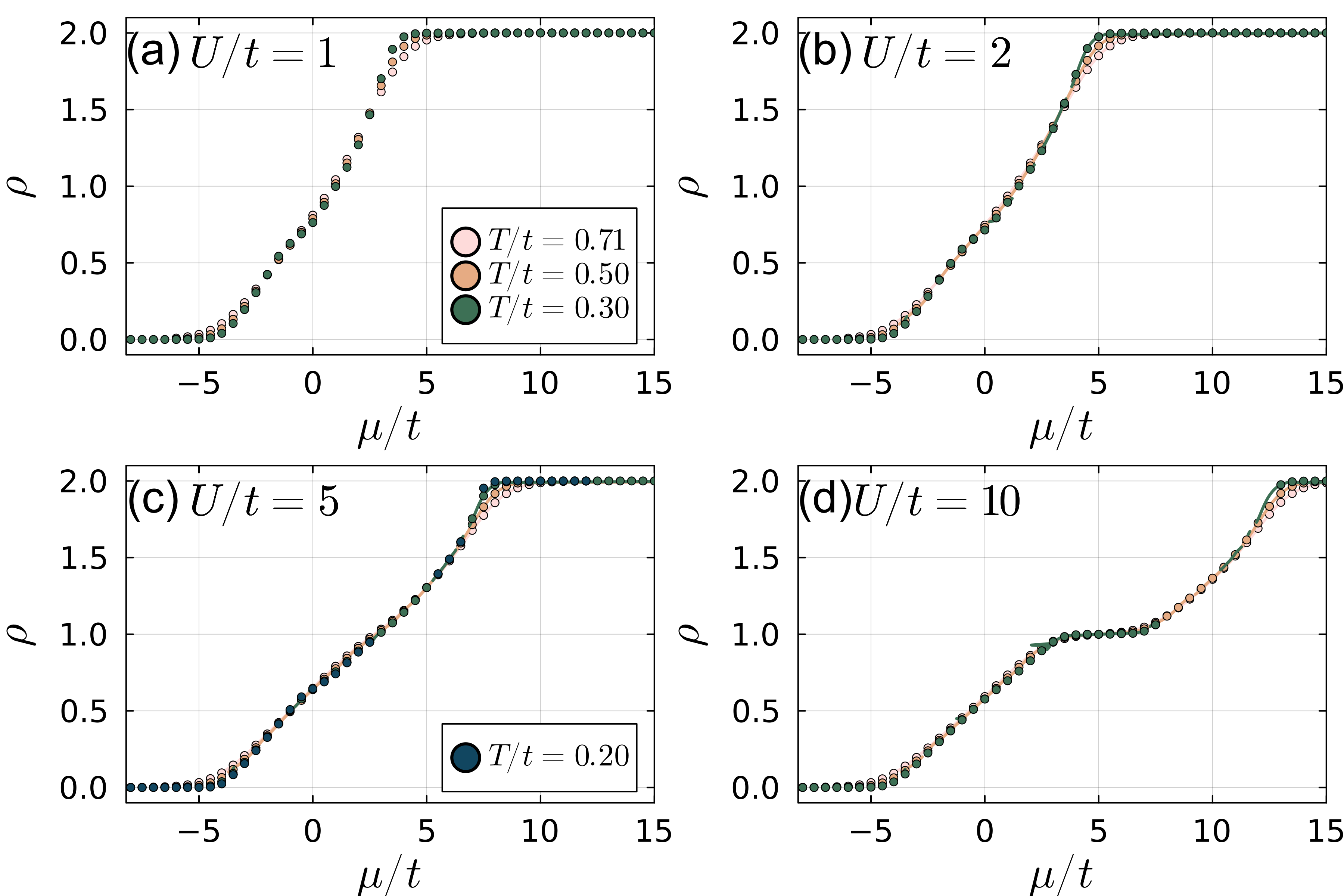}
    \caption{The equations of state for select $U/t$ values (a) $1$, (b) $2$, (c) $5$, and (d) $10$. Solid  lines are converged NLCE results in order 9 after Euler ressumation and circles are from the DQMC simulations. As $U/t$ increases to $10$, we see the development of the Mott plateau around half-filling.}
    \label{fig:EqnState}
\end{figure}

The noninteracting flat band of the Kagome lattice is expected to have a significant effect on the low-temperature physics of the Hubbard model at large densities, especially in the weakly-interacting region, i.e., $U/t\lesssim 6$, the noninteracting bandwidth. We observe signatures of the flat band in the equations of state in that region in Fig.~\ref{fig:EqnState}, where we plot the average density ($\rho$) vs the chemical potential for $U/t=1$, $2$, $5$, and $10$ and at several temperatures as low as $20\%$ of the hopping amplitude. When $U/t=1$ [Fig.~\ref{fig:EqnState}(a)], we find that curves at temperatures $T/t=0.71$ to $0.30$ cross at a point around $\rho=1.5$, which is inside the flat band of the noninteracting model. The density increases by decreasing the temperature at $\rho\gtrsim 1.5$ and decreases at $\rho\lesssim 1.5$, suggesting the development of a sharp rise in density at lower temperatures around that region, a direct sign of small dispersion in the many-body band~\cite{Kaufmann2021}. A similar trend can be seen for $U/t=2$ in Fig.~\ref{fig:EqnState}(b), and to some extent also for $U/t=5$ in Fig.~\ref{fig:EqnState}(c). We attribute these sharp features to remnants of the flat band.

At larger $U/t$ values, a Mott gap opens in the density of state at half filling, and the system undergoes a metal-insulator transition in the ground state. We observe the finite-temperature signature of the Mott gap for $U/t=10$ in Fig.~\ref{fig:EqnState}(d) as a flat region around $\mu=5$ with the density pinned at one. The size of the Mott gap will be proportional to $U$. The physics of the half-filled KLHM, including the value for the critical interaction at the metal-insulator transition, $U_c$, has been the subject of numerous numerical studies in the past~\cite{Bernhard_2007,Ohashi2006,Kuratani2007,PhysRevB.83.195127,Higa2016,Kaufmann2021,PhysRevB.104.L121118,Thereza2023}. In Appendix A, we discuss this transition and how we use our exact results for the compressibility in the thermodynamic limit to obtain an accurate $U_c/t=6.41(5)$, consistent with previous estimates based on DQMC studies which found $U_c/t=6.5 \pm 0.5$~\cite{Thereza2023} and $U_c/t=7.0$~\cite{Kaufmann2021}. The focus of this paper is to characterize the ferromagnetic correlations of the system that develop away from half filling, especially in the large density region near the band insulator.

\begin{figure}[t]
    \includegraphics[width=\linewidth]{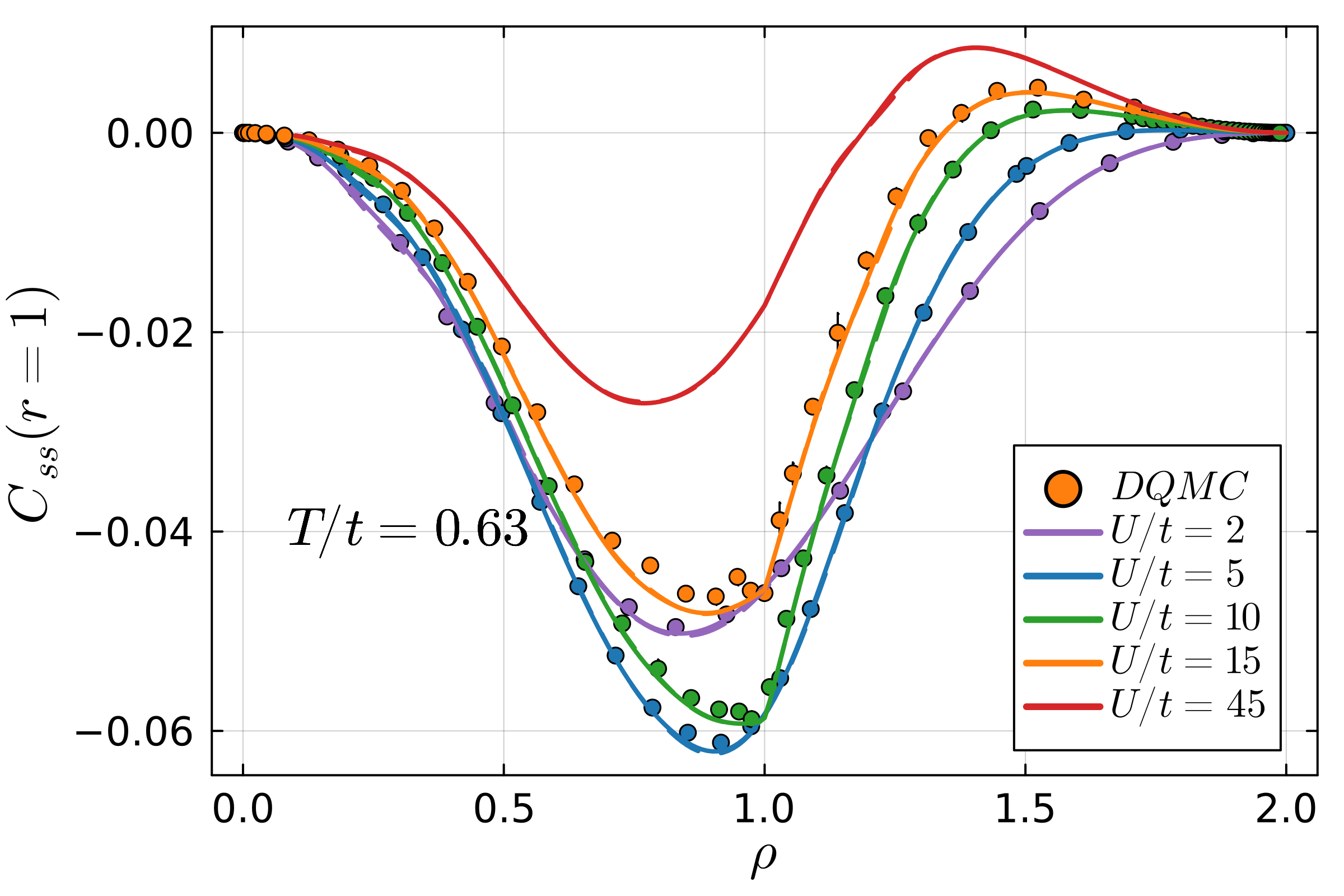}
    \caption{The nearest-neighbor spin correlation of the KLHM vs average particle density for different $U/t$ values at a fixed temperature of $T/t = 0.625$. Solid and dashed lines are NLCE results from the 9th and 8th orders of the expansion, respectively, after the Euler resummation. Circles are DQMC results.}
    \label{fig:CssVsRho}
\end{figure}

We find robust short-range ferromagneitc correlations at intermediate temperatures around the density of $1.5$ for large enough interaction strengths ($U/t\gtrsim 10$). Figure~\ref{fig:CssVsRho} shows spin correlations per bond,
\be
C_{ss}(r) = \braket{S^z_i S^z_{i+r}}-\braket{S^z_i}\braket{S^z_{i+r}},
\ee
where $S^z_i=\frac{1}{2}(n_{i\uparrow} - n_{i\downarrow})$, for the nearest-neighbor bonds ($r=1$) at $T/t= 0.625$. In the weakly-interacting region, they are antiferromagnetic at all densities at this temperature and largest in magnitude around half filling. Upon increasing $U/t$ from $2$ to $5$, they initially strengthen as more moments are formed, but are significantly suppressed as $U/t$ further increases beyond $5$. They even turn ferromagnetic (positive) at large densities, similarly to what has been observed for the triangular lattice model in recent optical lattice experiments~\cite{Xu2023,Lebrat2024}. Such ferromagnetic correlations are expected at finite temperatures above the ground state that is proven to be a ferromagnet on the Kagome geometry  for any nonzero $U$ at densities between $5/3$ and $11/6$ inside the flat band of the noninteracting system~\cite{Mielke_1992}.

Increasing $U/t$ from $\mathcal{O}(1)$ to $U/t\gg1$, however, will ultimately change the nature of the ferromagnetic ground state from a band-structure-driven one at large densities to that driven by Mott physics just above half filling, i.e., the Nagaoka ferromagnetism~\cite{y_nagaoka_65}. We already observe signatures of this change of character in nearest-neighbor correlations shown in Fig.~\ref{fig:CssVsRho}. For $U/t\sim 10$, the region with ferromagnetic nearest-neighbor correlations first appears around the particle doping of $\sim30\%$ at $T/t= 0.625$, which is at the edge of the noninteracting flat band in the density of states~\cite{yfwy-q6y9}. However, we find that this critical doping is largely dependent on $U$ and the temperature. Specifically, we find that for a large $U/t=45$ (red curve in Fig.~\ref{fig:CssVsRho}), not only the critical doping for ferromagnetic correlations has decreased significantly, but also the correlation function is approaching an antisymmetric form around half filling. This phenomenon, which was also observed for the triangular lattice~\cite{Lebrat2024}, can be understood from the fact that for $U/t\gg1$, the low-temperature physics of the system near half filling is dominated by a Mott insulator that can be insensitive to whether it is initially doped by particles or holes.

Here, the NLCE provides us with access to $U/t\gg1$, where the DQMC simulations become intractable. This highlights the complementarity of the two numerical methods, which otherwise show an excellent agreement in the common parameter regions for all the quantities we have studied in this paper. The temperature at which results in Fig.~\ref{fig:CssVsRho} are shown is chosen such that the NLCE results after resummations converge across all densities for all the $U$ values. However, the lowest convergence temperature for the NLCE, or the lowest temperature the DQMC can access before the sign problem becomes severe, depends on the density and the interaction strength. 

\begin{figure}[t]
    \includegraphics[width=\linewidth]{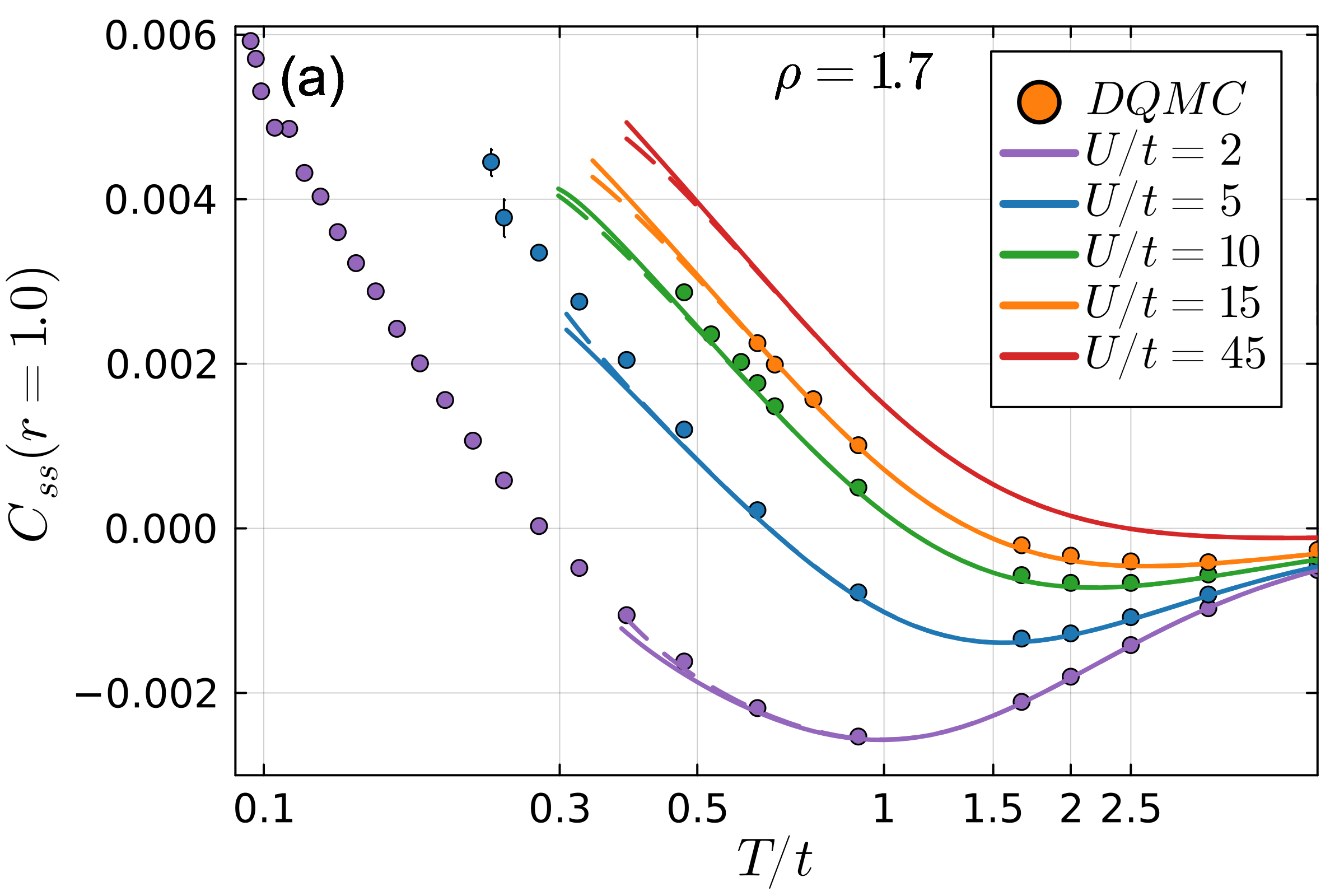}
    \includegraphics[width=\linewidth]{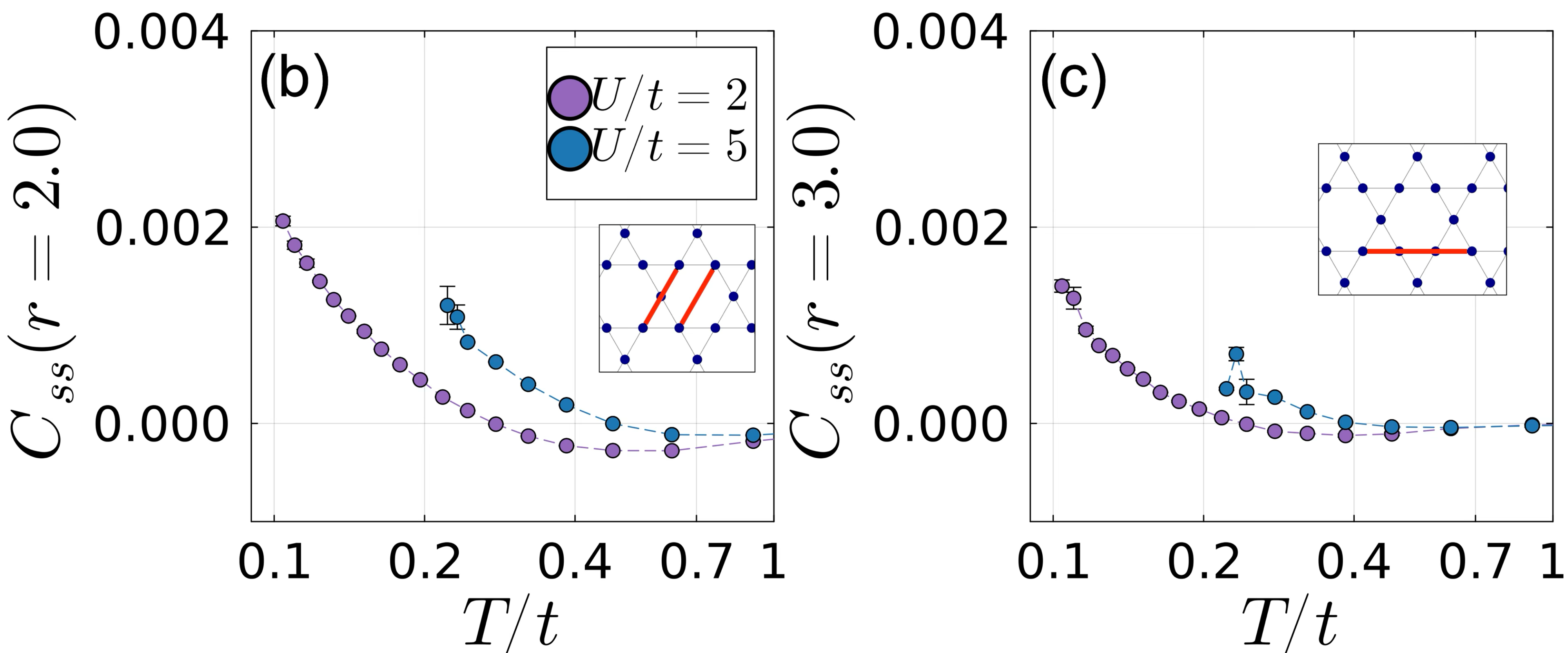}
    \caption{(a) The nearest-neighbor spin correlation of the KLHM  vs temperature for select $U/t$ values at a fixed average particle density of $\rho=1.7$. Lines and symbols are the same as in Fig.~\ref{fig:CssVsRho}. (b) \& (c) The same as in (a), except for longer-range correlations from DQMC at $r=2.0$ and $r=3.0$, respectively. Thin dashed lines are guides to the eye. The insets highlight the relevant bonds.}
    \label{fig:CssVsT}
\end{figure}

As the temperature is reduced, the system at large enough particle densities will develop FM correlations for any $U>0$, with the $U$-dependent crossover temperature (from AFM to FM) decreasing with decreasing the interaction strength. In Fig.~\ref{fig:CssVsT}(a), we show $C_{ss}(r=1)$ as a function of temperature for the same $U$ values as in Fig.~\ref{fig:CssVsRho} but at a fixed density of $\rho=1.7$ in the middle of Meilke's FM region. This gives us another perspective beyond that offered by Fig.~\ref{fig:CssVsRho} and makes the above observation clear. For example, even for the relatively small $U/t=2$, despite the AFM $C_{ss}$ initially strengthening upon decreasing the temperature, we find that $C_{ss}$ crosses zero to positive value around $T/t=0.3$ and continues to rise by lowering the temperature as seen in the DQMC data. For this $U$ value, DQMC can access temperatures as low as $T/t\sim 0.1$. We also observe that for $U/t=10$, the NLCE's convergence is extended to $T/t=0.3$ at this density.

\begin{figure}[t]
    \includegraphics[width=\linewidth]{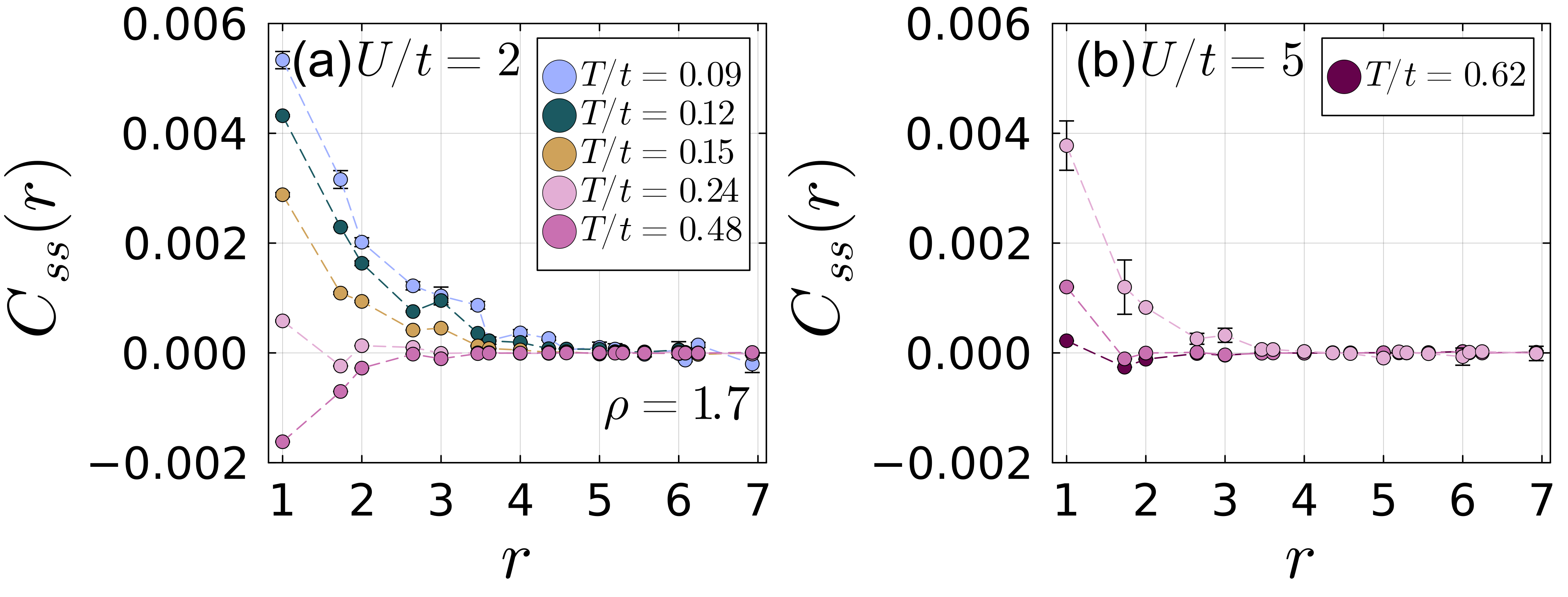}
    \caption{Spin correlations vs. distance from DQMC for (a) $U/t=2$ and (b) $U/t=5$ at different temperatures but all at a fixed average particle density of $\rho =1.7$. Lines are guides to the eye.}
    \label{fig:CssVsr}
\end{figure}

In order to gain better insight into the extent of ferromagnetic instability in the system, we have to look beyond nearest-neighbor correlations. In Figs.~\ref{fig:CssVsT}(b) and \ref{fig:CssVsT}(c), we show DQMC results for spin correlations vs temperature at the same density of $1.7$ but across longer distances of $r=2$ and $r=3$. At the lowest temperatures to which we have access, the correlations for $U/t=2$ turn ferromagnetic at both of these longer distances, signaling the development of long-range FM order. For $U/t=5$, while our access to correlations is limited to a higher temperature of $T/t=0.2$, we see a similar trend of FM correlations growing by lowering the temperature. 

These results are not specific to $r=2$ and $3$. Figure~\ref{fig:CssVsr} shows DQMC results for $C_{ss}(r)$ as a function of $r$ and at the same density of $1.7$. They confirm the development and extent of these FM correlations as the temperature decreases to $T/t=0.09$ for $U/t=2$ and to $T/t=0.24$ for $U/t=5$. 

\begin{figure}[t]
    \includegraphics[width=\linewidth]{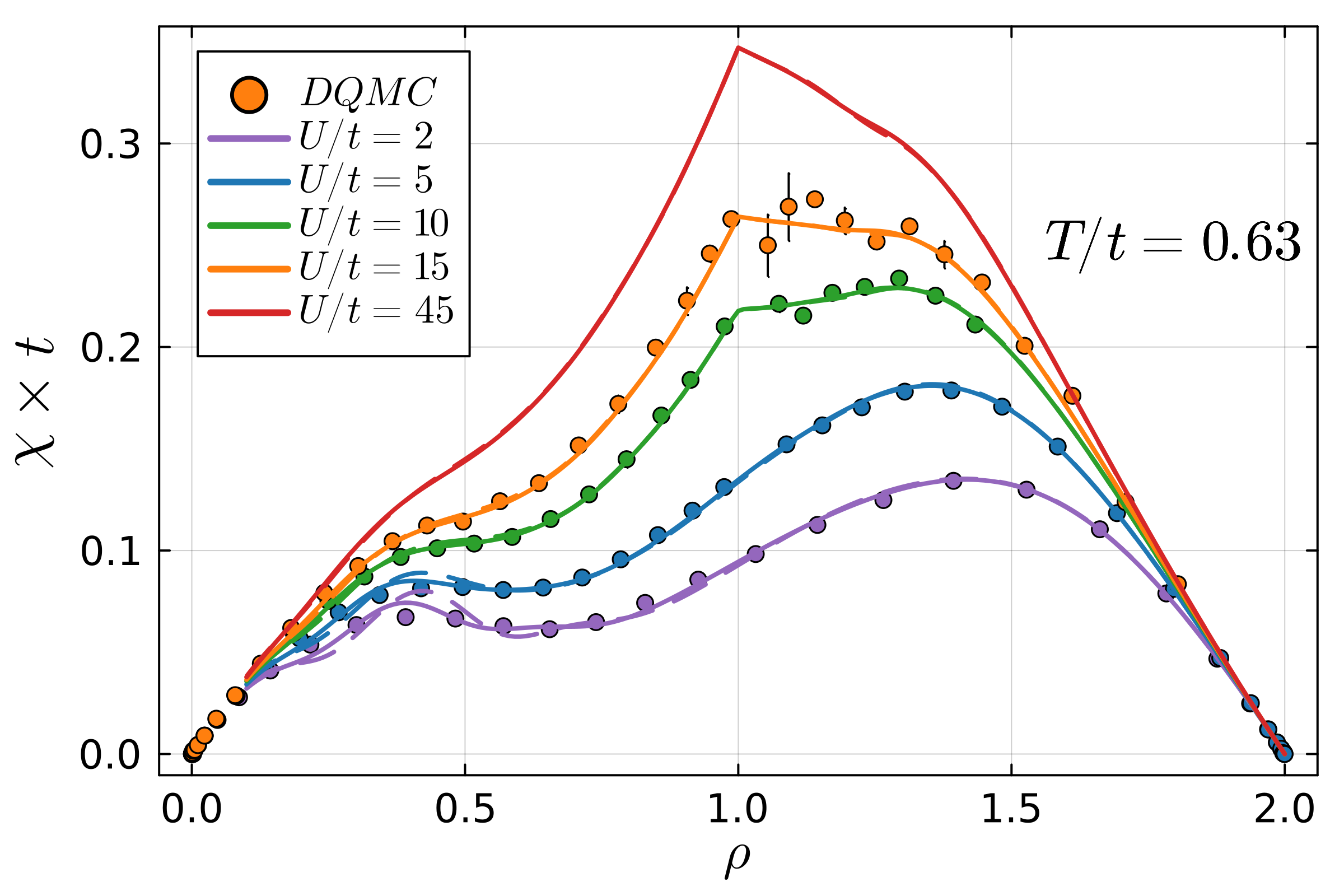}
    \caption{The magnetic susceptibility vs. average particle density at a fixed temperature $T/t = 0.625$ and for different $U/t$ values. Lines and symbols are the same as in Fig.~\ref{fig:CssVsRho}.}
    \label{fig:susVsRho}
\end{figure}

Next, we turn to the uniform magnetic susceptibility, $\chi$, which is equivalent to the FM structure factor divided by the temperature, and therefore, contains correlations at all length scales. It can also be easily obtained via fluctuations in magnetism: $\chi=(\left<M^2\right>-\left<M\right>^2)/T$, where $M=\braket{S^z_i}$. In Fig.~\ref{fig:susVsRho}, we show $\chi$ at the fixed temperature of $T/t=0.63$ as a function of density for several interaction strengths. For $U/t=2$, we find two broad peaks around fillings of $0.5$ and $1.5$. Upon increasing the interaction, roughly beyond $U_c$, we observe the development of a kink (discontinuity in the derivative), and eventually for $U/t>15$ a peak, at half filling at this temperature. The appearance of the kink for $U/t\ge 10$, which is the result of a sudden rise in the susceptibility at the commensurate filling, coincides with the opening of the charge gap for the same interaction values and temperature (see Fig.~\ref{fig:kappaVsRho} in the Appendix), and therefore, can be attributed to the sudden change in the availability of moments. However, due to the AFM nature of the correlations at $\rho\le 1$, we do not expect a divergence with further decreasing the temperature.

\begin{figure}[b]
    \includegraphics[width=\linewidth]{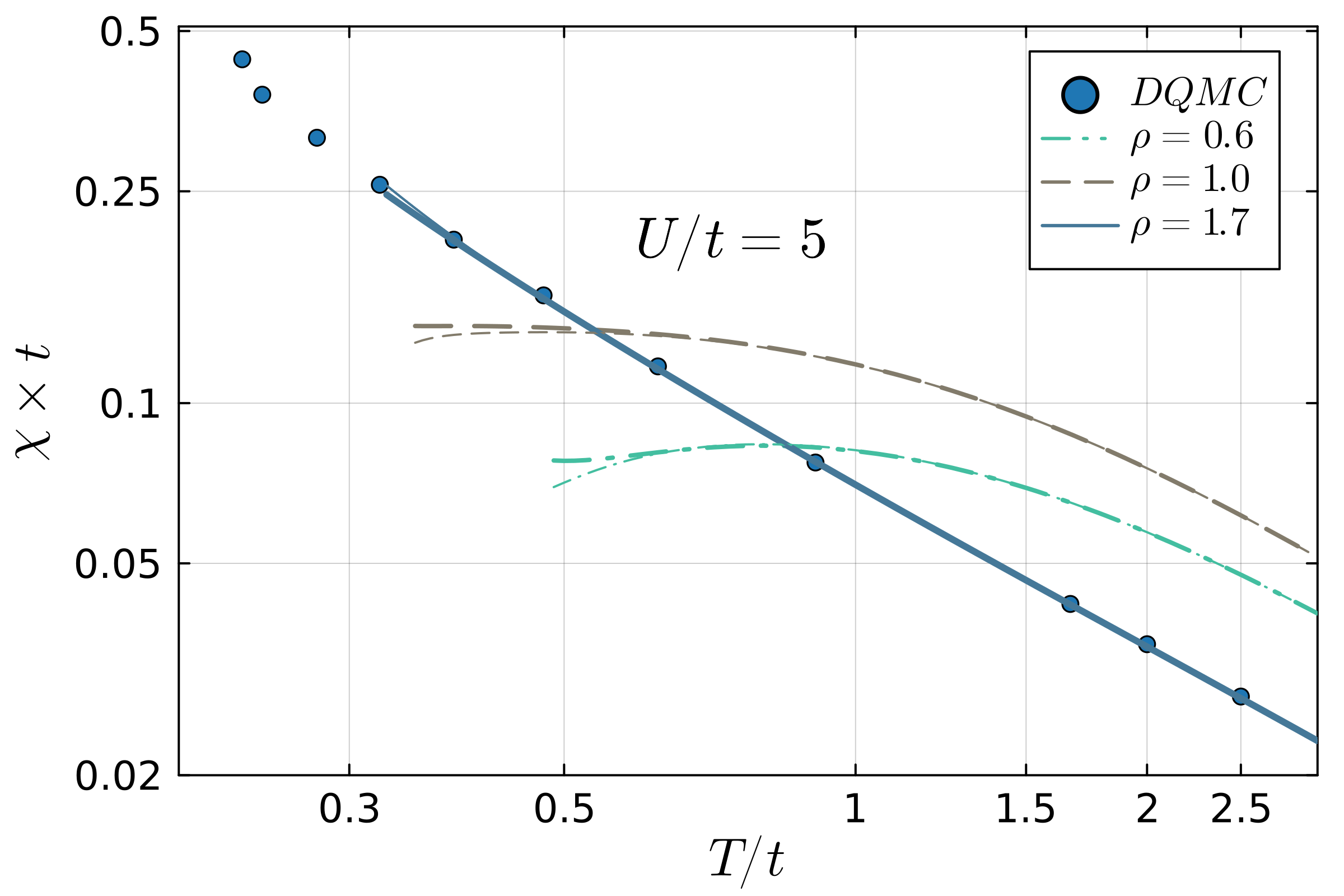}
    \caption{Log-log plot of magnetic susceptibility vs temperature for $U/t=5$ at particle densities $0.6,1.0$, and $1.7$. Thick (thin) lines are NLCE results from the 9th (8th) order of the expansion after resummation and symbols are DQMC results for $\rho=1.7$. The results suggest that this quantity peaks at intermediate temperatures when $\rho=0.6$ or $\rho=1.0$. For $\rho=1.7$, however, we observe a strong monotonic growth upon decreasing the temperature to $T/t\sim 0.2$.}
    \label{fig:magVsTU5}
\end{figure}

To gain more insight into the temperature trends in the susceptibility, in Fig.~\ref{fig:magVsTU5}, we plot the same quantity for $U/t=5$ as a function of temperature at select densities below, at, and above half filling. We find that NLCE's lowest convergence temperatures can extend below $T/t = 0.63$ for $\rho\gtrsim 1.7$, which together with DQMC results at $\rho=1.7$ at even lower temperatures make clear that FM correlations at $\rho=1.7$ continue to grow rapidly as the temperature is lowered while at half filling, they remain dominant only down to $T/t\sim0.5$ for this interaction strength. We find similarly diverging magnetic susceptibilities for all $U/t>0$ at large enough densities above half filling (not shown), consistent with Miekle's proof for the existence of FM near full filling~\cite{Mielke_1992}.

Although Mermin-Wagner theorem~\cite{M-W} conveys the lack of divergence of the susceptibility at finite temperatures in our two-dimensional quantum system, we find that its behavior at intermediate temperatures is consistent with the Curie-Weiss form for magnetic instability: $\chi=\frac{C}{T-\Theta_{CW}}$, where $C$ is a system-dependent constant and $\Theta_{CW}$ is the Curie-Weiss critical temperature. Therefore, we utilize this mean-field form to quantify the tendency of the system towards the FM instability. This technique has been previously adopted in the context of FM in the triangular lattice Hubbard model~\cite{Tang2020,Lee2023,Morera2023}, in which $\Theta_{CW}$ is extracted from a fit of the magnetic susceptibility at intermediate temperatures to the Curie-Weiss form and used to assess whether or not the ground state is likely ferromagnetic. Namely, a positive $\Theta_{CW}$ is taken as an indication for FM instability in the ground state while a negative $\Theta_{CW}$ is taken to mean such an instability is unlikely and that the ground state is AFM.

\begin{figure}[t]
    \includegraphics[width=\linewidth]{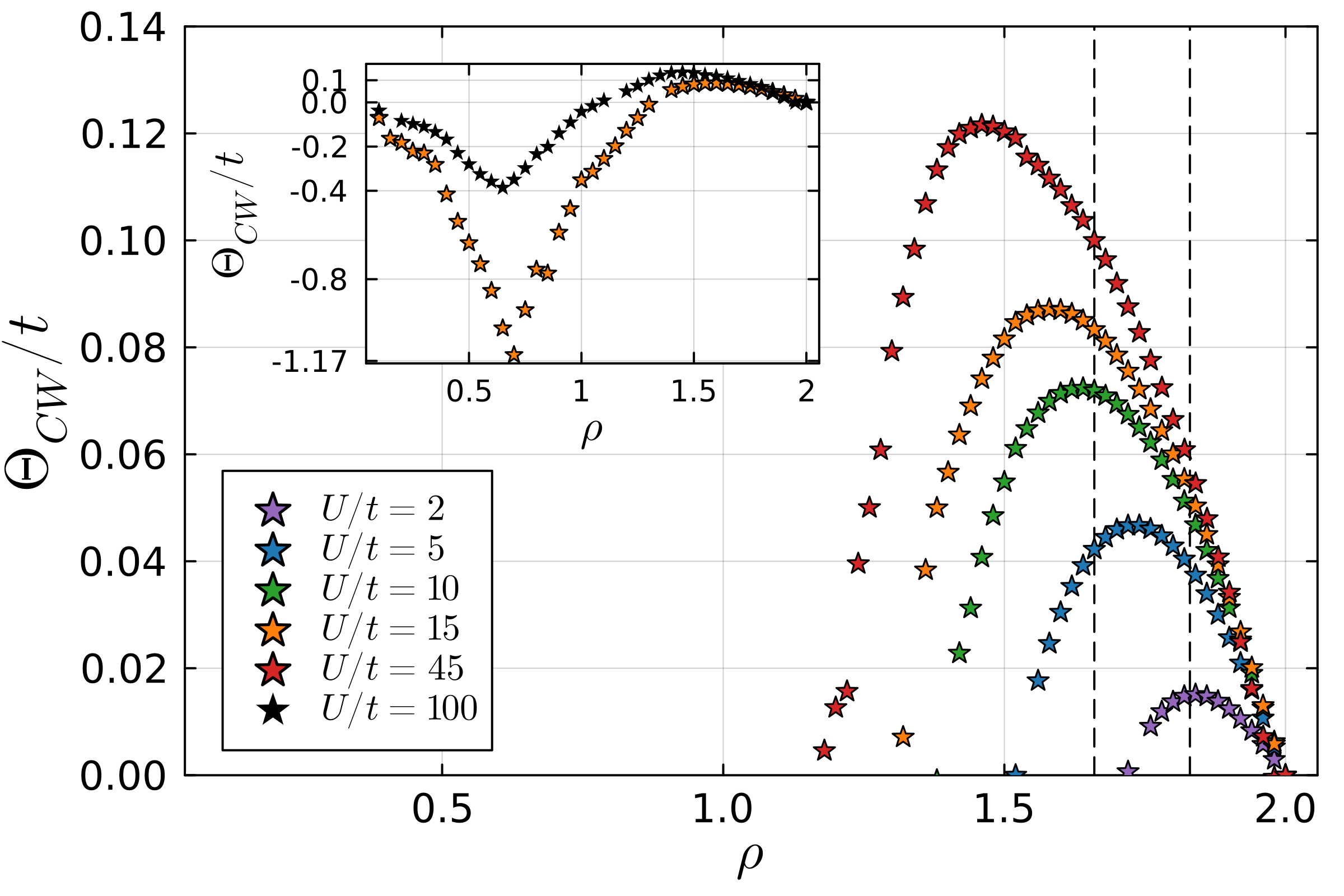}
    \caption{Curie-Weiss temperature, extracted from fits of NLCE data to $\chi=\frac{C}{T-\Theta_{CW}}$, vs particle density for multiple values of interaction strength $U/t$. Dotted black lines mark the region proven to be a ferromagnet in the ground state for any $U>0$~\cite{Mielke_1992}.}
    \label{fig:thetaVsRho}
\end{figure}

Following the same idea, we fit our magnetic susceptibility from the 9th order NLCE after the Euler resummation to the Curie-Weiss form in the temperature range $0.5<T/t<2.0$ (using $C$ and $\Theta_{CW}$ as the fitting parameters) and plot the resulting $\Theta_{CW}$ vs particle density in Fig \ref{fig:thetaVsRho}. For all $U/t$, we find a region with $\Theta_{CW}>0$ that starts at a $U$-dependent doping in the particle-doped side and extends to full filling. We find that as $U/t$ increases, this critical doping decreases and approaches half filling. We also observe that $\Theta_{CW}$ increases overall with positive values becoming larger and negative values becoming smaller (see the inset of Fig. \ref{fig:thetaVsRho}), although that does not necessarily mean stronger FM or AFM correlations in the ground state \cite{Lee2023}. As is also clear in the inset, the critical doping for $U/t=100$ is very close to half filling. These trends in $\Theta_{CW}$ are similar to those seen for $C_{ss}$ in Fig.~\ref{fig:CssVsRho} and are consistent with the ground state changing character from flat band FM in the weak-coupling region to Nagaoka FM in the strong-coupling region.

%%%%%%%%%%%%%%%%%%%%%%%%%%%%%%%%%%%%%%%%%%%%%%%%%%%%%%%%%%%%%%%%%%%%%%%%%%%%%%%
\section{Discussion}
\label{sec:discuss}
%%%%%%%%%%%%%%%%%%%%%%%%%%%%%%%%%%%%%%%%%%%%%%%%%%%%%%%%%%%%%%%%%%%%%%%%%%%%%%%

The KLHM is a notoriously difficult quantum lattice model to simulate numerically. In this work, we utilized two exact finite-temperature methods, the NLCE and the DQMC, to study its  properties in a full range of densities and a wide range of interaction strengths from near the noninteracting limit to extreme values, of the order of tens of the hopping amplitude. We focused on the magnetic correlations, which demonstrate the development of robust and long-range ferromagnetic correlations in the system on the particle-doped side. We found that the region with strong FM correlations starts at the band insulator and large densities inside the flat band when $U/t$ is small compared to the noninteracting bandwidth, and gradually extends towards half filling as $U/t$ increases to extreme values, consistent with Nagaoka FM. In the absence of any errors and given the validity of the NLCE results in the thermodynamic limit, we were also able to extract the charge gap at half filling from charge compressibility at intermediate temperatures and obtain, to the best of our knowledge, the most accurate estimate of the critical interaction strength for the Mott metal-insulator transition so far (see Appendix).

Our finite-temperature results for the KLHM can be used to guide future experiments utilizing optical lattices to study the model at lower temperatures to which we had access. Past efforts have demonstrated the possibility of realizing the Kagome geometry optically~\cite{PhysRevLett.108.045305,PhysRevA.101.011601,PhysRevLett.125.133001} and explored the effect of interactions between bosonic atoms loaded on the lattice on the flat band~\cite{PhysRevLett.125.133001}. Recent experiments have realized tunable geometries~\cite{Xu2023,BlochComm} and another multi-band model with a flat band, the Lieb lattice~\cite{lebrat2025}, (see also Ref.~\cite{lange2026} for a proposal to realize the Emory model) with interesting connections to the Kagome lattice~\cite{PhysRevB.108.235163}.

The ground state of the KLHM, especially away from half filling and in regions relevant to long-range FM order, is still an unexplored problem. Recent progress in the use of neural quantum states~\cite{Lange_2024} as ansatz for variational Monte Carlo studies of quantum lattice models, especially the Fermi-Hubbard model~\cite{doi:10.1073/pnas.2122059119,PhysRevResearch.7.013122,ByteDance,roth2025}, offers a promising tool to access the essential magnetic and superconducting~\cite{PhysRevB.105.075118} properties of the KLHM in the ground state on lattice sizes beyond the reach of other more conventional methods. An important question would be the extent of the FM phase at large particle dopings beyond Mielke's prediction, its dependence on $U$, and whether or not the nature of the FM phase changes as one approaches the Nagaoka limit. Another interesting problem concerns the hole-doped model and the existence and extent of Haerter-Shastry's kinetic AFM.

\acknowledgements

This work was supported by the grant DE-SC0022311 funded by the U.S. Department of Energy, Office of Science. F.C. acknowledges support from the National Science foundation's Research Traineeship (NRT) Fellowship funded by the Grant No. DGE-2125906. E.K. thanks Ivan Morera Navarro for insightful discussions about the Curie-Weiss temperature. E.K. is grateful to Juan Carrasquilla for hosting him at the Institut f\"{u}r Theoretische Physik at ETH Z\"{u}rich where a portion of the project was completed. E.K.’s contribution also benefited from his time as a Fellow at the Kavli Institute for Theoretical Physics (KITP), which is supported in part by the NSF grant PHY-2309135 and by the Heising-Simons Foundation. Computations were performed on the Spartan high-performance computing facility at San Jos\'{e} State University supported by the NSF under Grant No. OAC-1626645.

 \begin{figure}[t]
    \includegraphics[width=\linewidth]{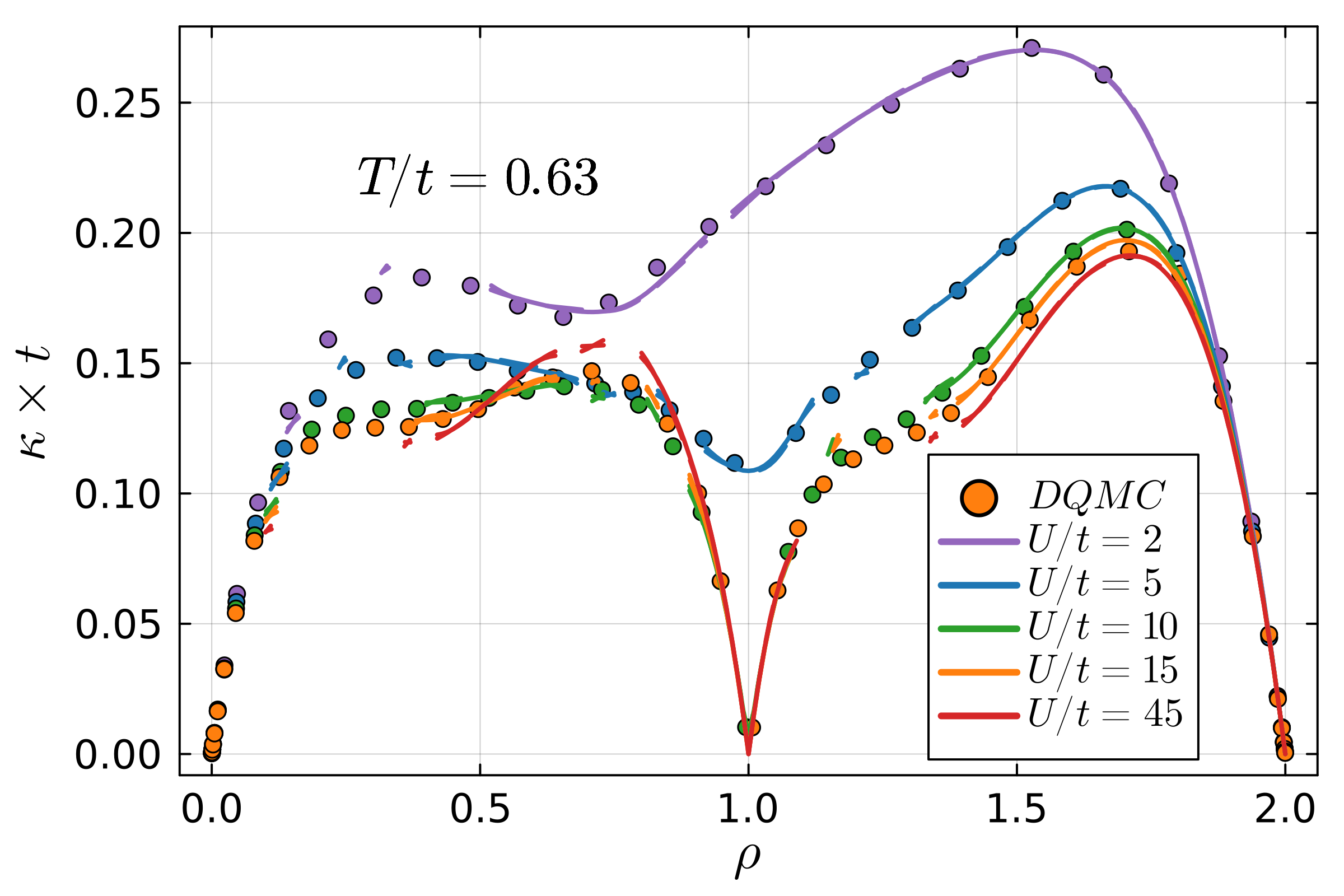}
    \caption{The compressibility vs density at a fixed temperature of $T/t=0.625$ and for different $U/t$ values. Lines and symbols are the same as in Fig.~\ref{fig:CssVsRho}. The compressibility at half filling decreases rapidly at this temperature upon increasing $U/t$ from $5$ to $10$, a finite-temperature signature of the Mott metal-insulator transition in the ground state.}
    \label{fig:kappaVsRho}
\end{figure}

\section*{Appendix}
\label{apndx}

\begin{figure}[t]
    \includegraphics[width=\linewidth]{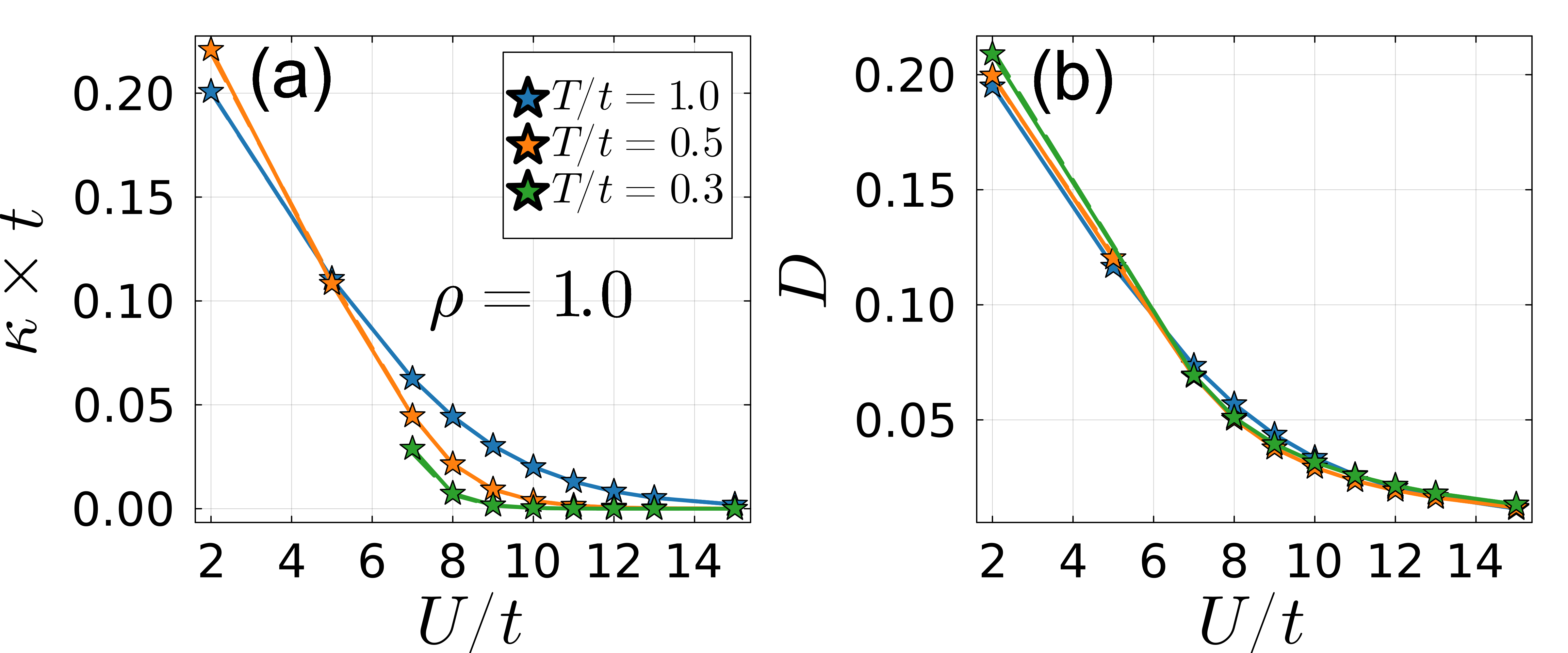}
    \caption{NLCE results for the (a) compressibility and (b) double occupancy vs $U/t$ at half filling at temperatures $T/t=1.0$, $0.5$, and $0.3$. We observe a sharp drop in $\kappa$ around $U\sim 6$. The double occupancy also shows a crossing of curves at different temperatures around the same interaction strength.}
    \label{fig:doubleAndKappaVsU}
\end{figure}

We find clear signs of Mott physics in our finite-temperature results upon increasing the interaction strength at half filling. In Fig.~\ref{fig:kappaVsRho}, we show the compressibility, $\kappa=\frac{\partial n}{\partial \mu}$ as a function of the density at $T/t=0.625$ for the same $U/t$ values as those used for the presentation of the magnetic properties in the main sections. As we increase the interaction strength from $U/t=5$ to $U/t=10$, $\kappa$ at half filling drops to values close to zero and remains very small upon further increasing $U/t$, signaling the strong suppression of charge degrees of freedom and the Mott physics setting in. Figure~\ref{fig:doubleAndKappaVsU} further shows the dependence of $\kappa$, as well as the average double occupancy $D=\braket{n_{i\uparrow} n_{i\downarrow}}$, at half filling on $U/t$ at fixed low temperatures. The plot of $\kappa$ in Fig.~\ref{fig:doubleAndKappaVsU}(a) displays the rapid decrease in this quantity with increasing the interaction strength around $U/t=7$, which becomes sharper as the temperature decreases. The double occupancy also shows a crossing of curves from different temperatures around $U/t=6$.

\begin{figure}[b]
    \includegraphics[width=\linewidth]{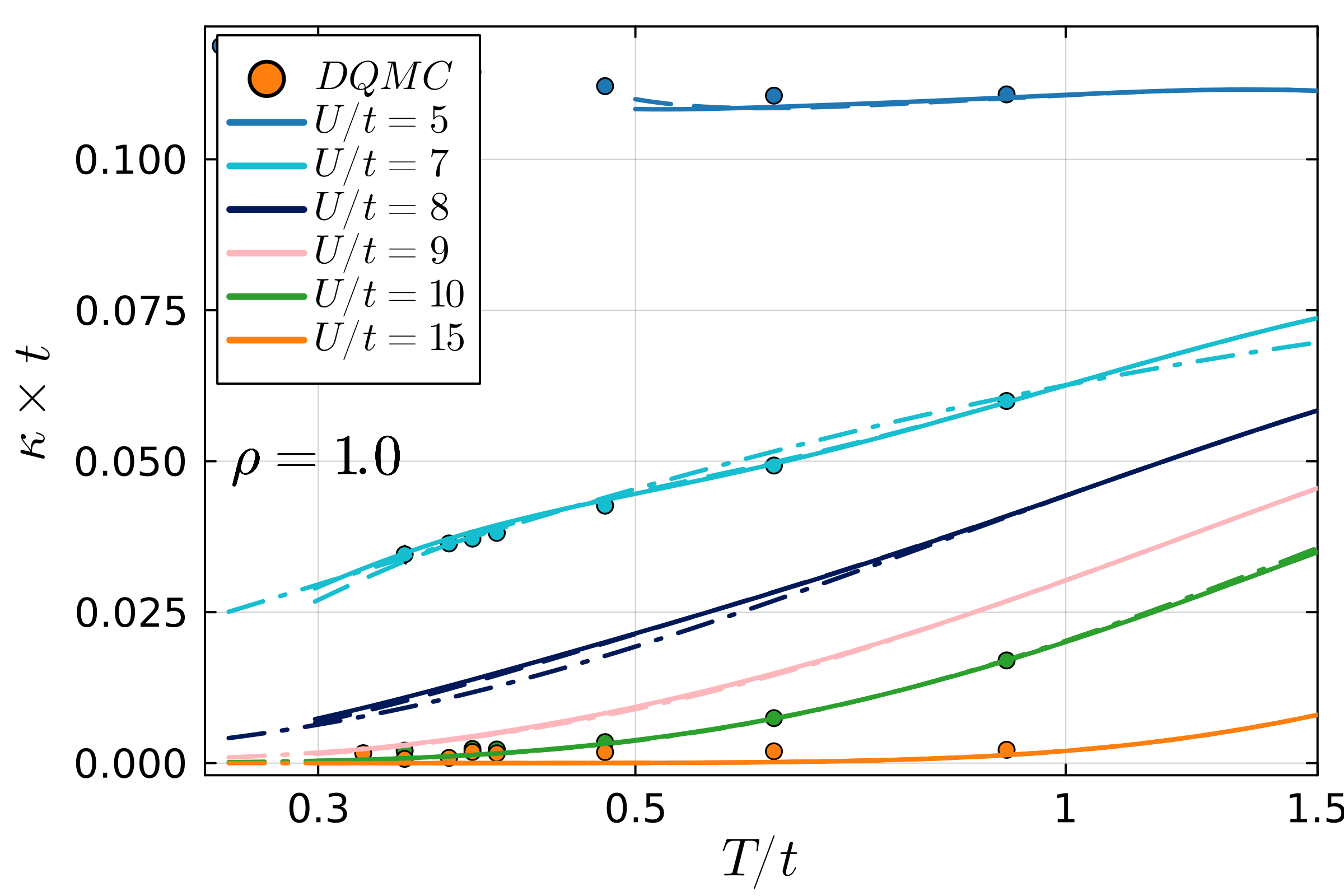}
    \caption{The compressibility of the KLHM vs temperature at half-filling for select $U/t$ values. Dash dotted lines are fits of the form $\kappa\times t(T)=a\exp(-\Delta_c/T)$.}
    \label{fig:kappaVsT}
\end{figure}

\begin{figure}[t]
    \includegraphics[width=\linewidth]{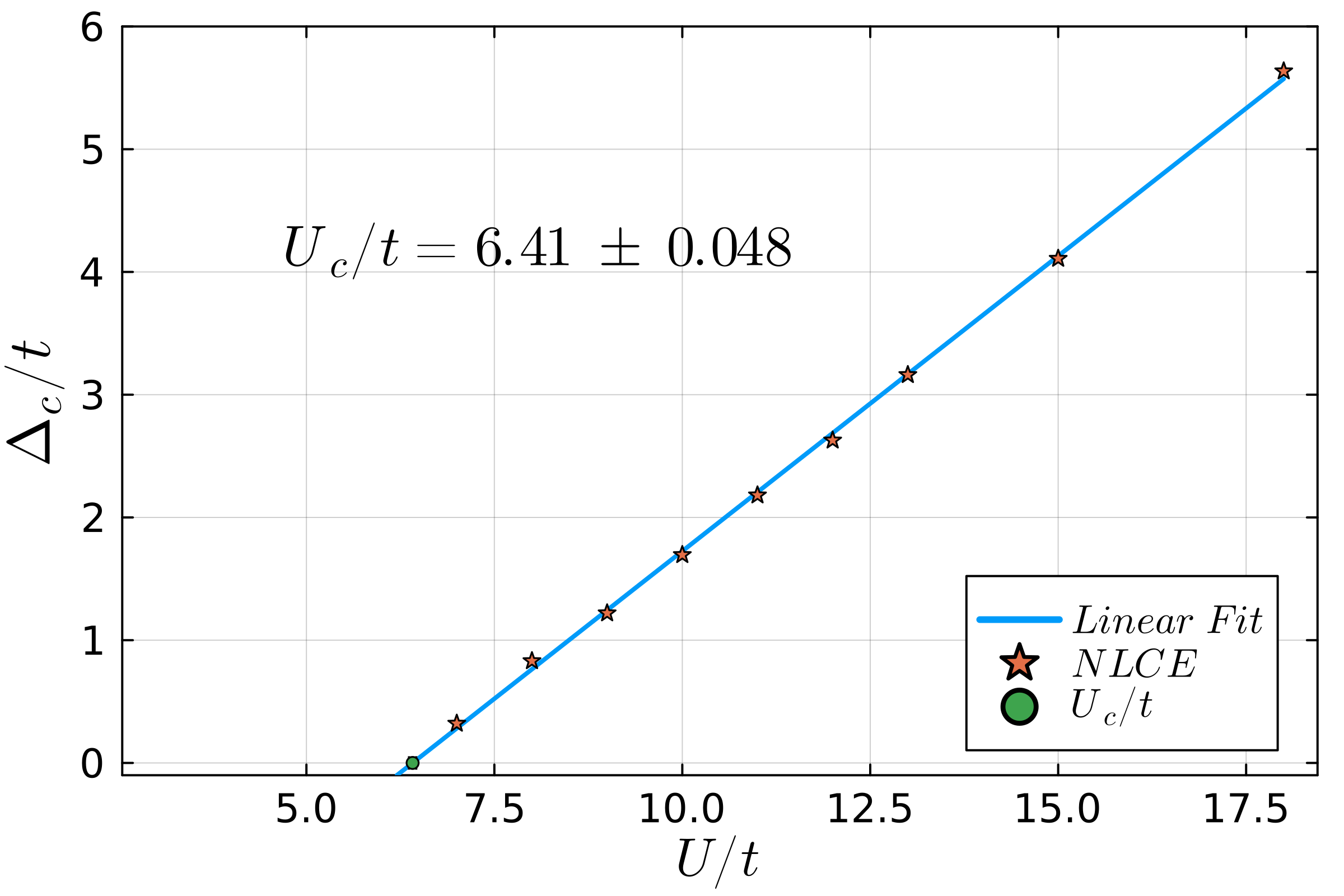}
    \caption{Charge gap $\Delta_c$, extracted from the fit of compressibility at half filling, versus interaction strength $U/t$. The blue line is a linear fit of the charge gap. 
    }
    \label{fig:delta}
\end{figure}

Using exact NLCE results for the compressibility at intermediate temperatures, we are able to obtain an accurate estimate for $U_c$ for the Mott metal-insulator transition in the ground state of the half-filled KLHM. In Fig.~\ref{fig:kappaVsT}, we plot $\kappa$ at half filling as a function of temperature for several values of $U$. We observe an activated behavior for $U/t\gtrsim 7$. Fitting the NLCE values between temperatures $T/t=0.3$ and $T/t=1.5$ to the form $\kappa\times t(T)=a\exp(-\Delta_c/T)$, where $\Delta_c$ is the charge gap and $a$ is another fitting parameter, we extract $\Delta_c$ for $U/t\ge 7$ and plot it as a function of $U$ in Fig.~\ref{fig:delta}. We observe a near perfect linear dependence of $\Delta_c$ on $U$, which leads us to $U_c/t=6.41(5)$ at the zero crossing. This critical value is consistent with our observations in Fig.~\ref{fig:EqnState} and the sharp features seen in $\chi$ and $\kappa$ at half filling when $U/t\ge 10$ in Figs.~\ref{fig:susVsRho} and \ref{fig:kappaVsRho}. To the best of our knowledge, this is the most accurate estimate of $U_c$ so far and is consistent with findings of previous numerical studies of the KLHM~\cite{Ohashi2006,Kuratani2007,PhysRevB.83.195127,Higa2016,Kaufmann2021,PhysRevB.104.L121118,Thereza2023}. It is closest to those obtained based on DQMC simulations ($U_c/t=6.5 \pm 0.5$ and $U_c/t=7.0$ \cite{Kaufmann2021,Thereza2023}).

\end{document}